\documentclass[pra,twocolumn,latterpaper,superscriptaddress, 10pt]{revtex4-1}  % for PRA using revtex4-1. Authors are separated by institution using groupedaddress in place of superscriptaddress
\usepackage{graphicx}  % needed for figures
\usepackage[usenames,dvipsnames]{color}
\usepackage{dcolumn}   % needed for some tables
\usepackage{bm}        % for math
\usepackage{amssymb}   % for math
\usepackage{amsmath,textcomp}
\usepackage{xfrac}
\usepackage{natbib}
\usepackage[utf8]{inputenc}
\usepackage{float}

\begin{document}

\title{Radiation-Pressure-Mediated Control of an Optomechanical Cavity}

\author{Jonathan Cripe}
%\email{jcripe1@lsu.edu}
\affiliation{Department of Physics \& Astronomy, Louisiana State University, Baton Rouge, LA, 70808}

\author{Nancy Aggarwal}
\affiliation{LIGO - Massachusetts Institute of Technology, Cambridge, MA 02139}

\author{Robinjeet Singh}
\affiliation{Department of Physics \& Astronomy, Louisiana State University, Baton Rouge, LA, 70808}

\author{Robert Lanza}
\affiliation{LIGO - Massachusetts Institute of Technology, Cambridge, MA 02139}

\author{Adam Libson}
\affiliation{LIGO - Massachusetts Institute of Technology, Cambridge, MA 02139}

\author{Min Jet Yap}
\affiliation{Australian National University, Canberra, Australian Capital Territory 0200, Australia}

\author{Garrett D. Cole}
\affiliation{Vienna Center for Quantum Science and Technology (VCQ), Faculty of Physics, University of Vienna, A-1090 Vienna, Austria}
\affiliation{Crystalline Mirror Solutions LLC and GmbH, Santa Barbara, CA, and Vienna, Austria}

\author{David E. McClelland}
\affiliation{Australian National University, Canberra, Australian Capital Territory 0200, Australia}

\author{Nergis Mavalvala}
\affiliation{LIGO - Massachusetts Institute of Technology, Cambridge, MA 02139}

\author{Thomas Corbitt}
\email{tcorbitt@phys.lsu.edu}
\affiliation{Department of Physics \& Astronomy, Louisiana State University, Baton Rouge, LA, 70808}

	\date{\today}
%\input author_list.tex       % D0 authors (remove the first 3 lines
                             % of this file prior to submission, they
                             % contain a time stamp for the authorlist)
                             % (includes institutions and visitors)
\date{\today}

\begin{abstract}
We describe and demonstrate a method to control a detuned movable-mirror Fabry-P\'erot cavity using
radiation pressure in the presence of a strong optical spring. At frequencies below the optical spring resonance, self-locking of the cavity is achieved intrinsically by the optomechanical (OM) interaction between the cavity field and the movable end mirror. The OM interaction results in a high rigidity and reduced susceptibility of the mirror to external forces.  However, due to a finite delay time in the cavity, this enhanced rigidity is accompanied by an anti-damping force, which destabilizes the cavity. The cavity is stabilized by applying external feedback in a frequency band around the optical spring resonance. The error signal is sensed in the amplitude quadrature of the transmitted beam with a photodetector. An amplitude modulator in the input path to the cavity modulates the light intensity to provide the stabilizing radiation pressure force.

\end{abstract}

%\pacs{}
\maketitle

\section{Introduction}
Cavity optomechanics, the interaction between electromagnetic radiation and mechanical motion, provides an ideal platform for measuring mechanical displacements and preparing and detecting mechanical resonators in the quantum regime \cite{24}. In a simple cavity-coupled optomechanical system, the mechanical oscillator is driven by the radiation pressure force exerted by the probing laser field. The fluctuations in the radiation pressure force due to power fluctuations modulate the motion of the mechanical oscillator, effectively changing the length of the cavity and modifying the resonance condition of the cavity. This leads to changes in the optical power circulating inside the cavity, thus cyclically modulating the radiation pressure force exerted on the mechanical oscillator. This feedback results in the optical spring effect.

The optical spring effect was first discussed for Fabry-Perot cavities by Braginsky \cite{Braginsky_1, Braginsky_2}. Braginsky \textit{et al.} \cite{Braginsky_3}, Buonanno and Chen \cite{Chen}, and Harms \textit{et al.} \cite{Harms} proposed using the optical bar and optical spring to enhance the sensitivity of gravitational wave detectors. Over the past two decades, many experiments have observed the optical spring in a variety of systems \cite{13, 15, 16, 17, 18, 19, 21, 22, 23, 24, Singh_PRL} and used it to optically cool mechanical resonators \cite{1, 2, 3, 4, 5, 6, 14, 20, Mow-Lowry}. Furthermore, proposals to increase the sensitivity of Michelson-type gravitational wave detectors using the optical spring effect have included adding a signal-recycling cavity \cite{Meers_1, Heinzel_1, 1-1}, using a detuned cavity to amplify the interferometric signal \cite{Verlot}, adding a signal-extraction cavity or resonant sideband extraction \cite{Mizuno, Heinzel_2}, and dynamically tuning the cavities to follow a gravitational wave chirp signal \cite{Meers_2, Simakov}. Signal-recycling and signal-extraction cavities have been used in the GEO 600 \cite{GEO} and  Advanced LIGO \cite{LIGO}  gravitational wave detectors, and are planned to be used in Advanced VIRGO \cite{VIRGO}, and KAGRA \cite{KAGRA}.
%\NA{either this text should add a sentence on role of optical spring physics in these proposals or we should remove it? (from "Furthermore, proposals to increase... " onwards)}

For a blue-detuned high-finesse optomechanical Fabry-P\'erot cavity in which the cavity's resonance frequency is less than the laser frequency, the system's effective mechanical resonance frequency is shifted to a higher frequency than the mechanical oscillator's eigenfrequency via the addition of the optical spring constant. This  leads to self-stabilization of the optomechanical system at frequencies below the optical spring frequency \cite{9}. At the optical spring frequency, however, the lag in optical response due to the round trip optical delay leads to a dominating anti-damping force that renders the system unstable \cite{2-2, 15, 16}. Such anti-damping forces normally require active feedback control to stabilize the optomechanical dynamics \cite{2-2, 15}. 

%Conventionally, a Pound-Drever-Hall (PDH) scheme is utilized to stabilize the cavity where the phase shift in the reflected carrier field is mixed with a local oscillator to extract the error signal \cite{Drever1983}. The extracted error signal is then nominally fed back to the piezoelectric device mounted to the stiff mirror of the cavity.  The PDH scheme is an effective technique for stabilizing the dynamics of a cavity in a close resonance condition. For an optomechanical dynamics driven far off resonance, however, the low slope of the error signal in the PDH scheme may lead to high technical closed loop noise.
Conventionally, detuned cavities are locked by using a simple ``side of fringe'' locking method. In this method the error signal is obtained from the slope of the cavity intensity profile on a transmission/reflection photodetector. This error signal is filtered and fed back to a piezoelectric actuator on the cavity mirror or to the frequency of the laser. The lock bandwidth is limited by the piezoelectric device's mechanical resonance frequency. The laser frequency modulation on the other hand has more bandwidth, but requires a large actuation range for short length cavities. 
%($L\delta \omega=\omega\delta L$)
As an alternative, in previous experiments, we have demonstrated the stabilization of the optomechanical cavity by utilizing the double optical spring effect \cite{Singh_PRL}. 

\begin{figure}
\center
\includegraphics[width=0.9\columnwidth]{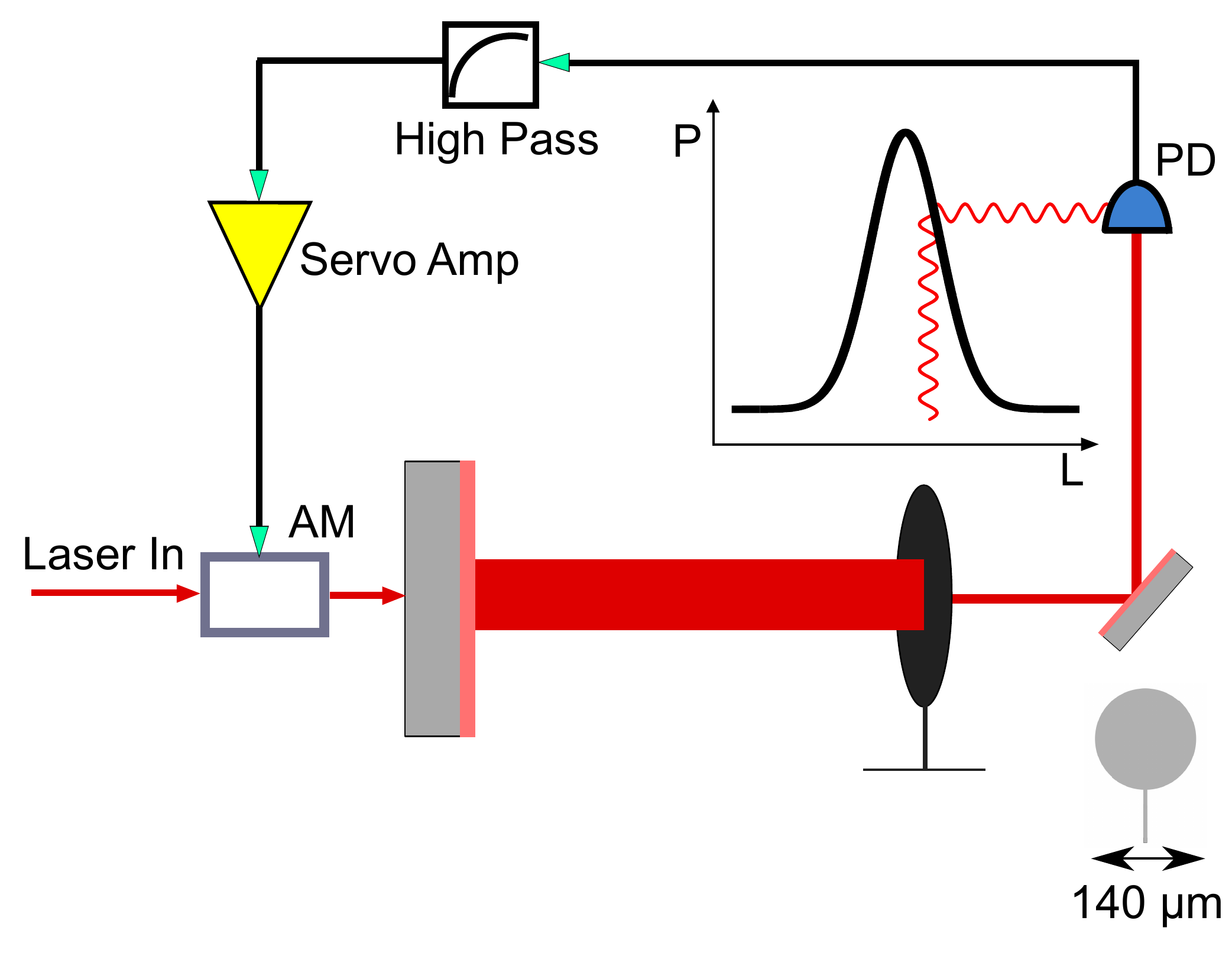}
\caption{Schematic of the experimental setup. The laser is passed through an amplitude modulator (AM), which modulates the intensity of the field before entering the detuned optomechanical cavity. The light is detected in transmission of the cavity and passed through a high pass filter and servo controller containing a gain and low pass filter component to obtain the error signal which is fed back to the AM. The inset shows the scale of the moveable mirror that forms one of the cavity mirrors.}
\label{Schematic}
\end{figure}

In this paper, we introduce a new feedback control method to lock a movable mirror Fabry-P\'erot cavity using radiation pressure. We have implemented this scheme at two independent experiments at LSU and MIT. This scheme relies on the suppression of external disturbances by having a large in-loop optomechanical gain as a result of the large optical spring constant. This suppression, which is mediated via the radiation pressure force, lowers the fluctuations in cavity length and power. A schematic representation of the method is shown in Fig. \ref{Schematic}, where the error signal is derived from the transmitted power out of the cavity and is used to control the radiation pressure force acting on the cavity by modulating the intensity of the input laser field passing through an amplitude modulator (AM). An optimal error signal is extracted by passing the transmitted field through a filter. This filter comprises of a gain and a band-pass component. The gain and low-pass filter of the servo controller are to stabilize the anti damping on the optical spring. The high-pass filter is to avoid saturation of the AM actuator due to ambient/seismic fluctuations that are largest at low frequencies (below a few kHz for a typical lab environment). These seismically and acoustically driven fluctuations in cavity length are self-stabilized in the optomechanical dynamics due to the high OM gain at frequencies below optical spring. 

\section{Theoretical Framework}
In understanding the noise stabilization of a strong optical spring system with feedback, it is informative to view the optical spring itself as a feedback mechanism. In this view, a closed-loop feedback system is formed between the mechanical oscillator and the optical cavity. The mechanical oscillator, with susceptibility $\chi_\mathrm{m}$, transduces a force into a displacement. The optical cavity, in turn, transduces the displacement back into a radiation pressure force, forming a closed loop. For simplicity, we consider the  frequency dependent susceptibility of a single mechanical resonance at $\Omega_\mathrm{m}$, such that
\begin{equation}
\chi_\mathrm{m} = \frac{1}{m\left(\Omega_\mathrm{m}^2-\Omega^2+i\Omega\Gamma_\mathrm{m}\right)},
\end{equation}
where $\Omega$ is frequency, $m$ is the effective mass of the mode of oscillation, $\Gamma_\mathrm{m} = \Omega_\mathrm{m} /Q_\mathrm{m}$ with $\Gamma_\mathrm{m}$ and $Q_\mathrm{m}$ the mechanical damping and quality factor of the mechanical oscillator, respectively. \cite{Note1}.
%\footnote{A more realistic form including multiple resonances could be used instead, but the single resonance susceptibility works well for this analysis. This is because the higher-order modes have a larger effective mass than the fundamental mode due to their poor overlap with the cavity mode and hence don't contribute much to the broadband behavior.}.

\begin{figure}
	\center
	\includegraphics[width= 0.95\columnwidth]{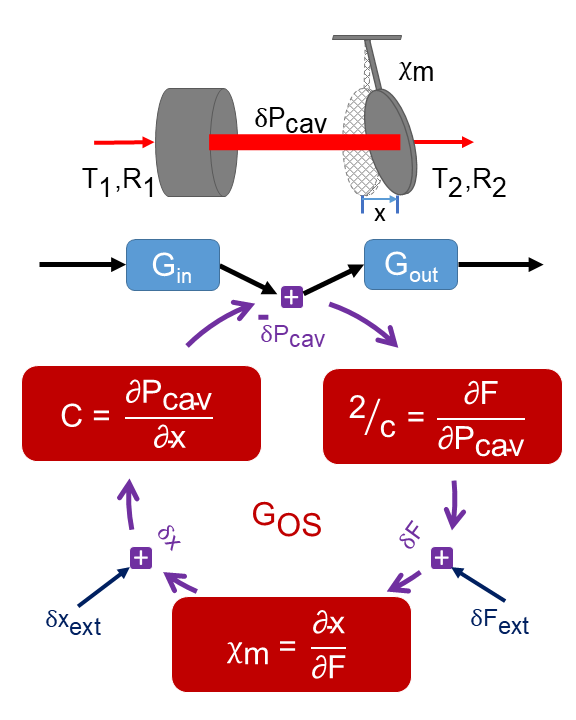}
	\caption{Detailed loop diagram for the cavity's transfer function. The amplitude modulator adds to the intra-cavity power $P_\mathrm{cav}$. The cavity power converts into radiation pressure force $F$, which then converts into cantilever displacement $x$ via its mechanical susceptibility $\chi_\mathrm{m}$. The displacement causes a length change for the cavity, leading again to a change in the intracavity power via the cavity response $C$. This forms a closed loop system.}     
	\label{fig:GOS}
\end{figure}

The open-loop gain pertaining to the cavity's closed-loop response as shown in Fig. \ref{fig:GOS} may be given in the limit $\Omega \ll \gamma$ as \cite{NAClosedLoopPaper}
\begin{eqnarray}\label{G_OS_Eq}
G_\mathrm{os}
&=& -\dfrac{32\pi \chi_\mathrm{m} P_\mathrm{cav}}{c\lambda_0 T} \dfrac {\delta_\gamma}{(1+\delta_\gamma^2)}  \left(1- \dfrac{2i\Omega}{\gamma(1+\delta_\gamma^2)}\right)\nonumber\\
&=& - \chi_\mathrm{m}K_0 \left(1-\dfrac{2i\Omega}{\gamma{[1+\delta_{\gamma}^2]}}\right)\nonumber\\
&=& - {m\chi_\mathrm{m}}\left(\Omega_{\mathrm{os}}^2 -i\Gamma_\mathrm{os}\Omega\right),\nonumber\\
&=&  \frac{\Omega_\mathrm{os}^2}{\left(\Omega_\mathrm{m}^2-\Omega^2+i\Omega\Gamma_\mathrm{m}\right)}\left(1-i\frac{\Gamma_\mathrm{os} \Omega}{\Omega_\mathrm{os}^2}\right)
\end{eqnarray}
where $P_\mathrm{cav}$ is the intra-cavity power, $\lambda_{o}$ is the center wavelength of the laser, $c$ is the speed of light, T is the total fraction of light leaving the cavity via loss and mirror  transmissions, $\gamma$ is the half width at half maximum (HWHM) for the cavity optical resonance in rad/s, $\delta_\gamma = \frac{\delta}{\gamma}=\frac{\omega_L-\omega_C}{\gamma}$ is the dimensionless detuning of the laser field from the cavity's resonance, and $K_0$ is the optical spring constant. The  optical spring frequency is given by  $m \Omega^2_\mathrm{os}=K_0$, and its HWHM is $\Gamma_\mathrm{os} =2 \Omega_\mathrm{os}^2/ \gamma / (1+\delta_\gamma^2)$. The real part of Eq. 2 corresponds to a position dependent restoring force and the imaginary part corresponds to a velocity dependent anti-damping force \cite{Note2}.
%\footnote{This force will be anti-restoring and damping for a red detuned laser.}.

\begin{figure}
	\center
	\includegraphics[width= 0.95\columnwidth]{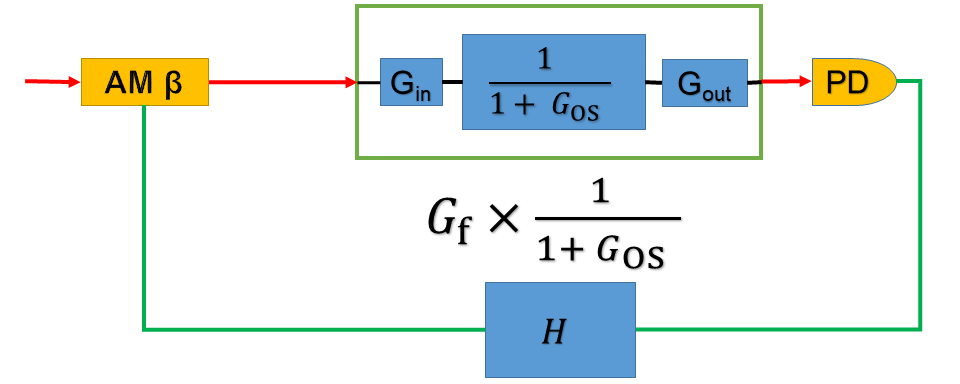}
	\caption{Loop diagram for the feedback $G_\mathrm{f}$. $H$ is the response of the high pass filter and servo controller, and $\beta$ is the response of the amplitude modulator. $\frac{1}{1+G_\mathrm{os}}$ is the closed loop response of the cavity's optical spring system, as shown in Fig \ref{fig:GOS}. $G_\mathrm{in}=\frac{4T_1 }{T^2}\frac{1}{1+\delta_\gamma^2}$ is the transfer function of the power input to power inside the cavity with the effect of the detuning taken into account, and similarly $G_\mathrm{out}=T_2$ is the transfer function from cavity power to transmitted power that is measured on the PD.
Here $T_1$ is the transmission of the input mirror, $T_2$ is the transmission of the microresonator, and $T$ is the total loss of power from the cavity, in the form of transmission, scattering, absorption, etc. $G_\mathrm{f}$ = $G_\mathrm{out} \times \mathrm{PD} \times H \times \beta \times G_\mathrm{in}$ is calculated by using the measurements of the individual transfer functions.}     
	\label{fig:loop_config2}
\end{figure}

 %The open loop gain magnitude crosses unity at about $\Omega = \Omega_\mathrm{os}$ under the approximation that $\Omega_\mathrm{os} \gg\Omega_\mathrm{m}$, but with negative phase margin $\tan^{-1}\left(\frac{\Gamma_\mathrm{os}}{\Omega_\mathrm{os}}\right)$. This negative phase margin must be stabilized by an external force. 
The effective susceptibility of the system to a force is then
\begin{eqnarray} \label{Chi_OS_Eq}
\chi_\mathrm{os}&=&\frac{x}{F_{\mathrm{ext}}}=\frac{\chi_\mathrm{m}}{1+G_\mathrm{os}} \nonumber\\
&\approx&\frac{1}{m}\frac{1}{\Omega_\mathrm{os}^2-\Omega^2 - i\Gamma_\mathrm{os}\Omega },
\end{eqnarray}
where $x$ is the displacement of the resonator, $F_{\mathrm{ext}}$, is an external force, and in the last step we assume that the $\Omega_\mathrm{os} \gg \Omega_\mathrm{m}$ and $\Gamma_\mathrm{os} \gg \Gamma_\mathrm{m}$. At frequencies below the optical spring frequency the ambient motion is therefore reduced by the factor
\begin{eqnarray} \label{Chi_m_OS_Eq}
\left| \frac{\chi_\mathrm{m}}{\chi_\mathrm{os}} \right| \approx \left|\frac{\Omega_\mathrm{os}^2}{\Omega_\mathrm{m}^2-\Omega^2+i\Omega\Gamma_\mathrm{m}}\right|, 
\end{eqnarray}
with the approximation assuming $\Omega_\mathrm{os} \gg \Omega$. This factor of suppression may be made very large if $\Omega_\mathrm{os}\gg \Omega_\mathrm{m} $. 

In the limit of a large optical spring frequency, the optical spring provides sufficient stabilization to maintain cavity lock. Due to the negative damping (gain) of the optical spring feedback, however, the system is unstable on its own. This can be seen by writing the closed-loop gain in the s-domain by substituting $s=i\Omega$,. 
The closed-loop gain $G_{\rm cl}$ corresponding to this open-loop gain $G_{\rm os}$ is given by 
\begin{eqnarray} \label{G_cl_Eq}
G_{\rm cl} &=& \frac{1}{1+G_{\rm os}}\nonumber \\
&=& \frac{\Omega_{\rm m}^2+s^2+s\Gamma_\mathrm{m}}{\Omega_{\rm m}^2+s^2+s\Gamma_\mathrm{m}+\Omega_\mathrm{os}^2-s\Gamma_{\rm os}}
\end{eqnarray}
From the above expression, one can see that this closed-loop gain has at least one right-half-plane pole \cite{Note3}
%\footnote{the terms in the denominator are not all of the same sign}
and will thus be unstable. This system must be stabilized by an external damping force. The feedback may be localized to frequencies near the optical spring resonance, and its only purpose is to stabilize the unity gain crossing.

The main purpose of the applied feedback $G_\mathrm{f}$ shown in Fig. \ref{fig:loop_config2} is to change the shape of the phase response of the system so that the system is stable as well as has good stability margins.

Radiation pressure is a natural transducer to stabilize such a system because there is strong coupling of radiation pressure by assumption. In addition, amplitude modulators have higher response bandwidth than piezoelectric actuators and better range than laser frequency modulation. Furthermore, because these systems are typically operated detuned (within a few line-widths to achieve strong optical springs), the transmitted power through the cavity is a natural readout of the cavity motion.

%\begin{figure}
%	%\begin{figure*}[h!]
%	\center
%	%\includegraphics[width= 0.9\textwidth]{layout5.pdf}
%	\includegraphics[width= 0.9\columnwidth]{layout5.pdf}
%	\caption{Configuration of feedback loop. The image in the center shows the full feedback loop with $T_2$ the transmission of the moveable end mirror, $H$ the response of the high pass filter and servo amp, and AM the response of the amplitude modulator. The mechanical susceptibility $\chi_\mathrm{m}$ transduces a force into a displacement. The cavity response C converts the displacement into a change in cavity power. The factor $2/c$ converts intracavity power back into force via radiation pressure. The diagrams to the left and right show the individual $G_\mathrm{os}$ and $G_\mathrm{f}$ feedback loops, respectively.}     
%	\label{fig:loop_config}
%	%\end{figure*}
%\end{figure}

\section{Experimental Setup}

The schematic shown in Fig. \ref{Schematic} illustrates the experimental setup. The laser field from an NPRO Nd:YAG laser is passed through an amplitude modulator before passing through a half-wave plate and mode-matching lenses en route to the optomechanical cavity. The in-vacuum cavity is 1 cm long and consists of a  0.5-inch (12mm) diameter input mirror with a 1 cm radius of curvature  and a microresonator as the second mirror. The input mirror is mounted on a piezoelectric actuator to allow for fine-tuning of the cavity length. The microresonator is fabricated from a stack of crystalline Al$_{ 0.92}$Ga$_{0.08}$As/GaAs layers. It has a diameter of 140 \textmu m and a mass of about 500 ng \cite{cole08, cole12, cole13, cole14, Singh_PRL}. The microresonator has a natural mechanical frequency of $\Omega_\mathrm{m}=2\pi\times288$ Hz and a measured mechanical quality factor $Q_\mathrm{m}=8000$, which gives $\Gamma_\mathrm{m} = 2\pi \times 36$ mHz. 

The field transmitted through the cavity is 
%split by a polarizing beam splitter with the transmitted beam
detected by a photodetector. The photodetector signal is sent through a high-pass filter and servo controller before being used as the error signal to the amplitude modulator. 

\section{Results and Discussion}

To help understand the feedback mechanisms and individual components of the feedback loops, Fig. \ref{fig:GOS} and Fig. \ref{fig:loop_config2} show the loop diagrams for the feedback loop $G_\mathrm{f}$ and the optical spring $G_\mathrm{os}$. Measurements of the open loop gain, plant transfer function, individual loop gains, and closed loop gain are described below.
\begin{figure}
	\center
	\includegraphics[width= \columnwidth]{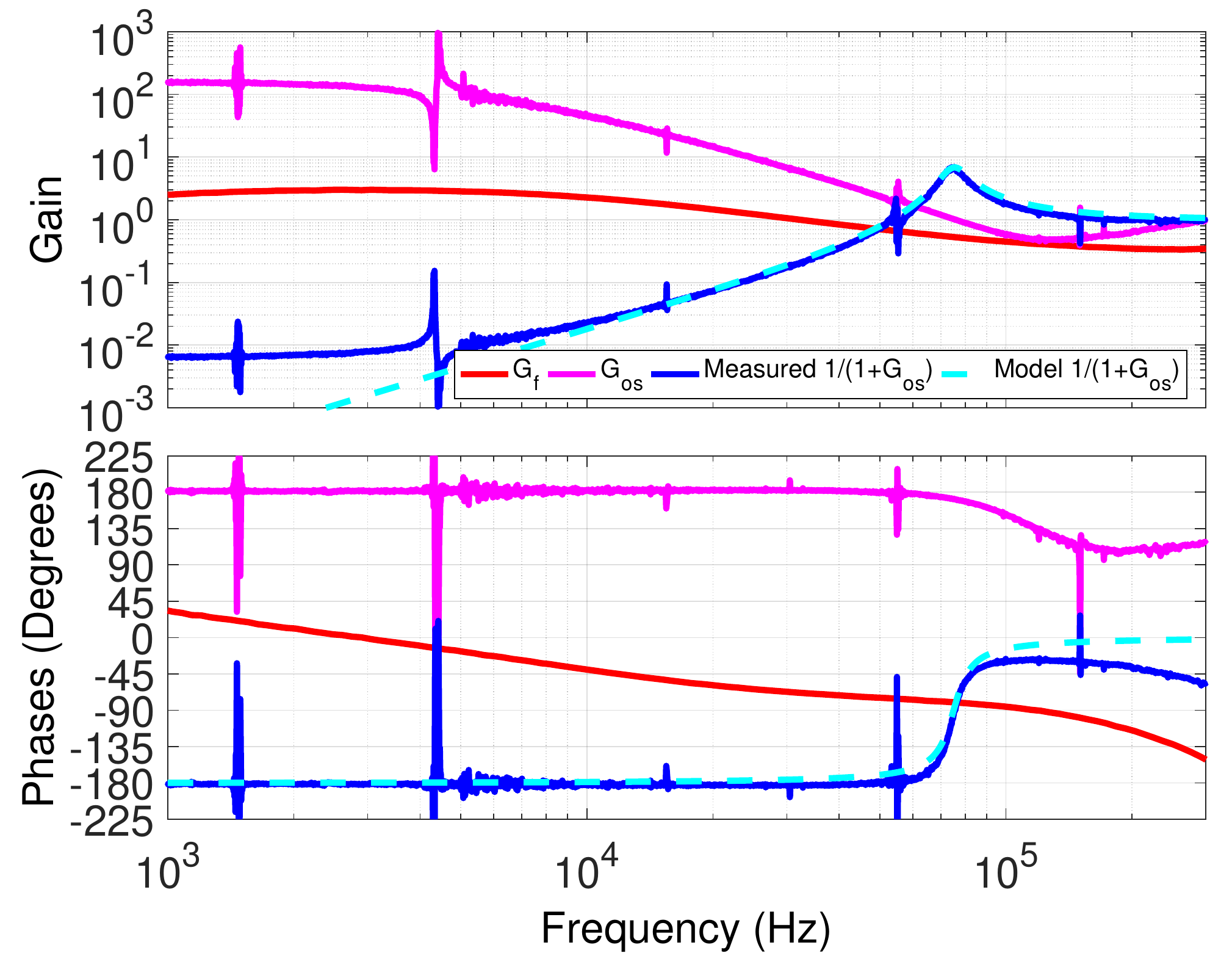}
	\caption{Transfer function measurements of the plant, $G_\mathrm{CL}=\frac{1}{1+G_\mathrm{os}}$ shown in blue, the feedback, $G_\mathrm{f}$ shown in red, and the open loop optical spring, $G_\mathrm{os}$ shown in pink. A model for the plant, shown in dashed cyan, is calculated with Eq. \ref{G_OS_Eq} using the measured values for $\Omega_\mathrm{os}$, $\Omega_\mathrm{m}$, and $\Gamma_\mathrm{m}$ and setting $\Gamma_\mathrm{os}$ so that the peak height and width match the measured data. $G_\mathrm{cl}$ is obtained using the open loop gain, $\frac{G_\mathrm{f}}{1+G_\mathrm{os}}$, shown in Fig. \ref{fig:Gf_1+Gos} and dividing out the measured $G_\mathrm{f}$. The effect of the optical spring is visible with the peak at 75 kHz and a rise in phase of the plant transfer function. The measurement begins to flatten out below 5 kHz due to other circuitous signal couplings (eg scattered light). $G_\mathrm{os}$ is then obtained from $G_\mathrm{cl}$. The applied feedback loop is $G_\mathrm{f}$. At 75 kHz, $G_\mathrm{f}$ has a magnitude of 0.53 and a phase of $-80^{\circ}$.}
	\label{fig:Gf_Gos_1_1+Gos}
\end{figure}

\begin{figure}
	\center
	\includegraphics[width=\columnwidth]{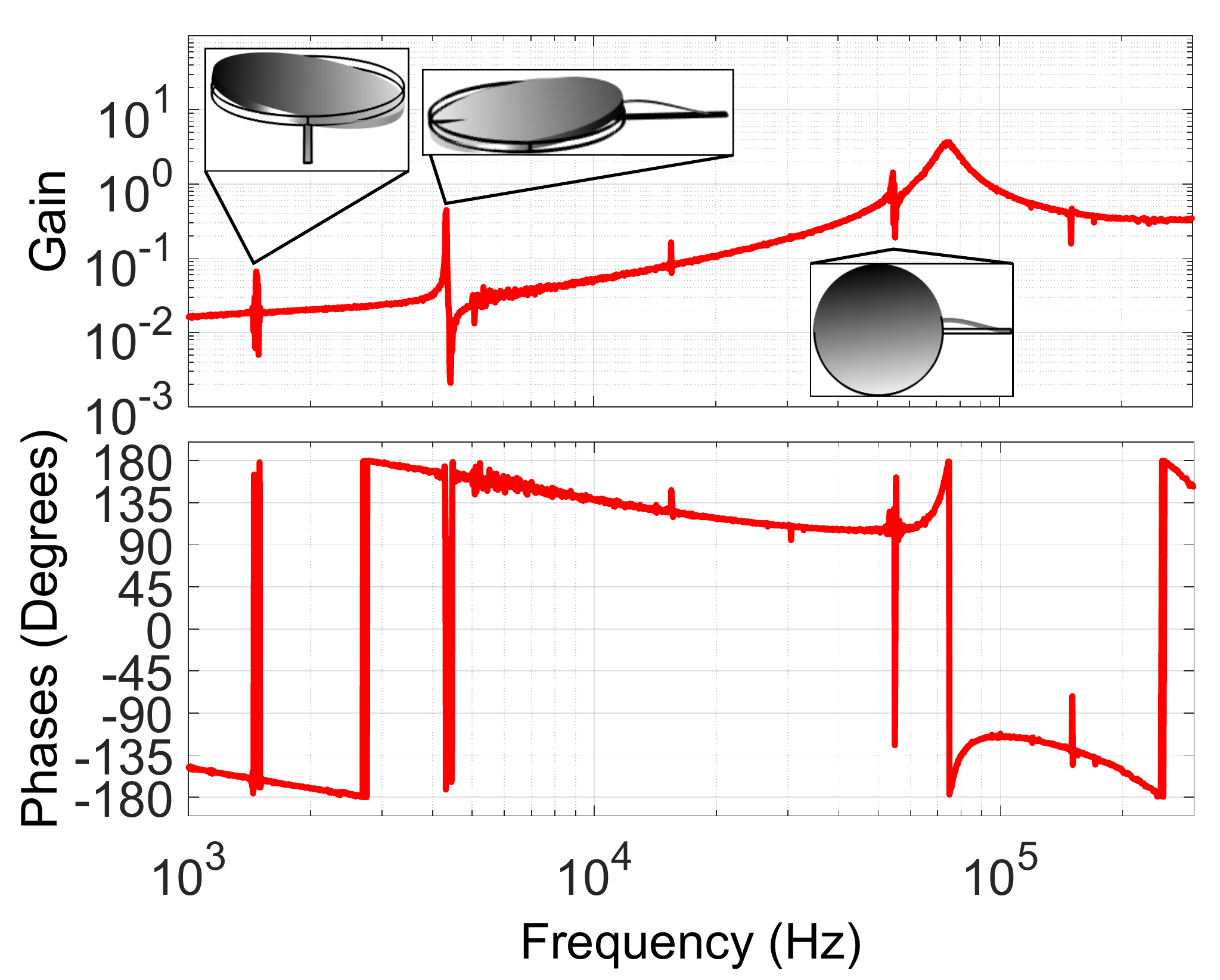}
	
	\caption{Measurement of the open-loop gain or $\frac{G_\mathrm{f}}{1+G_\mathrm{os}}$ taken by injecting a signal before $H$, done with a circulating power of 0.2 W. Higher-order mechanical modes are visible at 1.4 kHz (yaw), 4.3 kHz (pitch), and 54 kHz (translation and yaw) are shown in the inset. Unity-gain crossings are at 61 kHz and 93 kHz with phases  $109^{\circ}$ and $-115^{\circ}$. The gain is 0.34 at 250 kHz where the phase crosses $-180^{\circ}$. Thus, the system is stable with phase margins of $71^{\circ}$ and $65^{\circ}$, respectively, and a gain margin of 9.4 dB.}     
	\label{fig:Gf_1+Gos}
\end{figure}
In Fig. \ref{fig:Gf_Gos_1_1+Gos}, the blue curves show the plant transfer function, which is the system we would like to control. We see a peak corresponding to the optical spring at around 75 kHz in the magnitude. Since the system is unstable on its own, the plant transfer function is obtained using the open loop gain measurement with the feedback on. This open loop gain is shown in Fig. \ref{fig:Gf_1+Gos}, and we later divide it by the measured $G_\mathrm{f}$, shown in red in Fig. \ref{fig:Gf_Gos_1_1+Gos} to obtain the plant transfer function.

Fig. \ref{fig:Gf_Gos_1_1+Gos} also shows the transfer function of the $G_\mathrm{os}$ loop, which is obtained from the open loop transfer function shown in Fig. \ref{fig:Gf_1+Gos}. %$G_\mathrm{os}$ has a unity gain crossing at the optical spring frequency of 75 kHz. The phase at the unity gain frequency is $171^{\circ}$ or $-189^{\circ}$. This negative phase margin means that the feedback is positive instead of negative, which will cause the system to be unstable and oscillate out of control.
The large magnitude of $G_{\rm os}$ at frequencies below the optical spring shows the large suppression that the system's internal response is providing. 

The external electronic feedback loop, $G_\mathrm{f}$,  which is used to stabilize the system, is shown in Fig. \ref{fig:Gf_Gos_1_1+Gos} in the red curves. The measurement of $G_\mathrm{f}$ is obtained by measuring the response of individual elements in the loop, which includes the photodetector (PD), the high-pass filter and servo controller ($H$), and the amplitude modulator ($\mathrm{AM}$), and multiplying them together. The high-pass filter has a corner frequency at 800 Hz and the servo controller has a P-I corner at 100 kHz with a low frequency gain limit of 20 dB. We chose these values to supply sufficient phase margin while also attenuating the feedback at low-frequencies to avoid saturating the AM actuator. The measurement of the elements of $G_\mathrm{f}$ is done without using the cavity, so it gives the correct shape of $G_\mathrm{f}$, but does not provide the absolute scaling of the loop because the effect of the cavity is not included. The calibrated $G_\mathrm{f}$ is obtained by taking the effect of the cavity into account using the open-loop gain measurement above the optical spring peak where $\frac{1}{1+G_\mathrm{os}} \approx 1$.
%The calibrated $G_\mathrm{f}$ is obtained by taking the effect of the cavity, $G_{\rm in}$ into account using the open loop gain measurement.
%$G_\mathrm{in}$ and $G_\mathrm{out}$ into account from independently estimating the mirror reflectivities and cavity detuning.
%\agr{The scaling is obtained by looking at the $G_\mathrm{f}/(1+G_\mathrm{os})$ measurement above the optical spring frequency, where $G_\mathrm{os}$ is small, as shown in Fig. \ref{fig:Gf_Gos_1_1+Gos}. Therefore, $G_\mathrm{f}/(1+G_\mathrm{os}) \approx G_\mathrm{f}$. \agr{Gos is between 1 and 10 above the OS. Need to discuss.} The scaling factor is calculated by taking the ratio of the measurement of $G_\mathrm{f}/(1+G_\mathrm{os})$ above the optical spring and the measurement of $G_\mathrm{f}$. The calibrated $G_\mathrm{f}$ measurement is then obtained by multiplying the measured $G_\mathrm{f}$ by the scaling factor.}

Fig. \ref{fig:Gf_1+Gos} shows the measurement of the open loop gain taken by injecting a signal before $H$ (Fig. \ref{fig:loop_config2}) and measuring the response after PD. Since the measurement enters the $G_\mathrm{os}$ loop, the open loop transfer function is given by $\frac{G_\mathrm{f}}{1+G_\mathrm{os}}$. The effect of the optical spring is also visible in Fig. \ref{fig:Gf_1+Gos} with a resonance peak at 75 kHz and a falloff with $f^\mathrm{2}$ below the optical spring. There are two unity-gain crossings at 61 kHz and 93 kHz with phases of $109^{\circ}$ and $-115^{\circ}$. The gain at 250 kHz where the phase crosses $-180^{\circ}$ has a magnitude of 0.34. Thus, the system is stable with phase margins of $71^{\circ}$ and $65^{\circ}$, respectively, and a gain margin of 9.4 dB. We note that while the $G_\mathrm{f}$ shown in Fig. \ref{fig:Gf_Gos_1_1+Gos} does produce a stable system, it is not a unique solution. While other solutions for $G_\mathrm{f}$ may be more stable, the $G_\mathrm{f}$ we use is simple and achieves our goal of stabilizing the system. We also note that the measurement of $G_\mathrm{os}$ deviates from the expected $f^\mathrm{-2}$ slope above $\sim$ 100 kHz. This is a result of imperfect measurements of the individual components of the loops, which leads to errors in the subtraction for the transfer functions of $G_\mathrm{os}$ and $\frac{1}{1+G_\mathrm{os}}$.

\begin{figure}
	\center
	\includegraphics[width=\columnwidth]{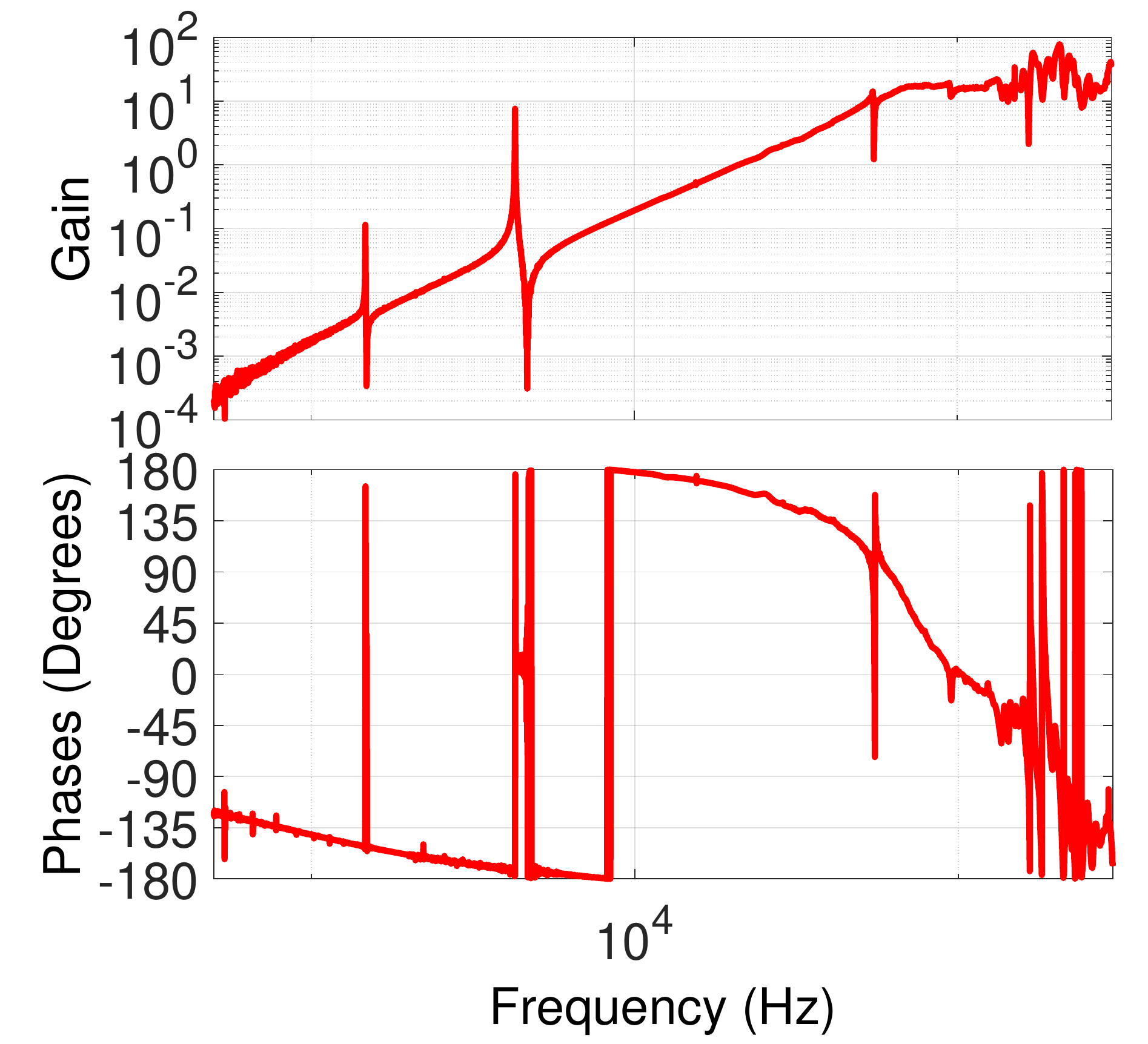}
	\caption{Measurement of the closed-loop response performed by modulating the laser frequency. This plot shows the suppression of low-frequency frequency noise below the optical spring frequency at $\approx 75$ kHz. The amount of suppression is calculated by taking the ratio of the measurement above the optical spring and at a low frequency below the optical spring. Using the values at 100 kHz and 500 Hz, noise is suppressed by a factor of at least 50,000. Higher order mechanical modes are again visible at 1.4 kHz, 4.3 kHz, and 61 kHz.}     
	\label{fig:Laser_freq_scan_Robi_table_combined}
\end{figure}
Another result of the dynamics of the optomechanical system is the reduced response to disturbances at frequencies below the optical spring frequency. Ambient motion causes the cavity length to change by $\Delta L \sim 10^{-7}$m, while the cavity linewdith is $\Delta \lambda \sim 10^{-11}$m. It is therefore necessary to suppress the ambient motion in order to operate the cavity. Fig. \ref{fig:Gf_Gos_1_1+Gos} shows the optical spring resonance at 75 kHz. According to Eq. \ref{Chi_m_OS_Eq}, the ambient motion should be reduced at low frequencies by the factor $\left|\frac{\Omega_\mathrm{os}^2}{\Omega_\mathrm{m}^2-\Omega^2+i\Omega\Gamma_\mathrm{m}}\right|$.

To verify this calculation, we modulate the laser frequency, which in effect, is the same as introducing a disturbance $\delta x_{\rm ext}$ in Fig. \ref{fig:GOS}. Fig. \ref{fig:Laser_freq_scan_Robi_table_combined} shows a measurement of 
%\begin{equation} \label{LFS_Eq}
%\frac{C \times \mathrm{PD} \times (1+G_\mathrm{f})}{1+ G_\mathrm{os}+G_\mathrm{f}}
%\end{equation}
\begin{equation} \label{LFS_Eq}
-\frac{L_p}{f_0}\frac{G_{\rm out} \times C \times PD}{1+G_{\rm f}+G_{\rm os}}
\end{equation}
taken by modulating the laser frequency and measuring the output of the PD with $\frac{L_p}{f_0}$ the change in the laser frequency for a given change in length for the laser piezo. The amount that low-frequency vibrations are reduced by is calculated by taking the ratio of the value of the measurement above the optical spring frequency where the measurement is flat and the value of the measurement at low frequencies, yielding a suppression of at least 50,000.

%\agr{Although in this plot I don't go down that low. Should I just use the numbers form this plot or should I somehow include the data from 200 Hz. I could include that as an inset? But I get 600,000 when I use that data, which is larger than we expect. I get 81,000 when I look at the data at 500 Hz. There is a dip at the mech res. 200 Hz is still part of the dip, so that could be why we have a disagreement. Need a theory plot for X/F to reference.}

The response of the system to an external force is
\begin{eqnarray} \label{X_F_Eq}
\frac{x}{F_ \mathrm{ext}} &=&  \frac{\chi_\mathrm{m} (1+G_\mathrm{f})}{1+ G_ \mathrm{os}+G_\mathrm{f}}.
\end{eqnarray}

According to Eq. \ref{X_F_Eq}, ambient fluctuations are suppressed by the factor $1+ G_ \mathrm{os}+G_\mathrm{f}$. Since this factor is in common in Eq. \ref{LFS_Eq} and Eq. \ref{X_F_Eq}, the laser frequency scan shown in Fig. \ref{fig:Laser_freq_scan_Robi_table_combined} is an accurate measure of the suppression of ambient fluctuations.

\section{Conclusion}

In conclusion, we have demonstrated a stable feedback control method to lock a moveable mirror Fabry-P\'erot cavity using radiation pressure. In this scheme, the use of radiation pressure as an actuator provides a large locking bandwidth compared to a piezoelectric device used in the simple ``side of fringe'' locking. We have experimentally shown that the system is stable and reduces low-frequency disturbances by a factor of at least 50,000. The combination of the stable system and excellent low-frequency noise suppression allows the optomechanical cavity to be operated on time scales of hours to days without losing lock. With the low-frequency noise reduced, we aim to measure broadband quantum radiation pressure noise and ponderomotive squeezing at frequencies relevant to Advanced LIGO. In addition, since the quadrature of the field inside the cavity is actually rotated with respect to the input field, the feedback gain could be increased by modulating in a different quadrature. A modulation in an arbitrary quadrature can be achieved by stitching together two amplitude modulator crystals and adding a relative drive between them \cite{UTM}. This configuration could be useful if the negative damping is too high to be compensated with a single amplitude modulator.

\begin{acknowledgments}
This work was supported by the National Science Foundation CAREER grant PHY-1150531, as well as PHY-1707840 and PHY-1404245. MY and DM receive support from the Australian Research Council through project number CE170100004 (OzGrav). This document has been assigned the LIGO document number LIGO-P1700100. NA would like to thank Slawomir Gras and Aaron Buikema for valuable discussions and suggestions.
\end{acknowledgments}

%\bibliographystyle{ieeetr} % Was previously using this bibliography style
%\bibliography{biblioRPL}

\end{document}